\documentclass[11pt]{article}
\usepackage{graphicx}

\begin{document}

\begin{center}
{\bf \Large Influence of a small fraction of individuals with enhanced 
mutations on a population genetic pool} 
\end{center}
\bigskip

\centerline{S. Cebrat and D. Stauffer$^*$}

\bigskip
\noindent
Department of Genomics, 
Wroc{\l}aw University, ul. Przybyszewskiego 63/77, 51-148 Wroc{\l}aw, Poland

\medskip
\noindent
$^*$ Visiting from Institute for Theoretical Physics, Cologne University, 
D-50923 K\"oln, Euroland

\begin{abstract}
Computer simulations of the Penna ageing model suggest that already a small
fraction of births with enhanced number of new mutations can negatively 
influence the whole population.
\end{abstract}
\section{Introduction}

Medical progress has prolonged enormously the life expectation in 
the industrialised world, compared with two centuries ago. But 
the case of antibiotics has told us also that there are disadvantages:
While many human lives were saved, many bacteria became resistant, and
new antibiotics have to be developed. In normal biological evolution,
this balance of effort and countermeasures is called the ``Red Queen'' 
effect. 

In particular for humans, computer simulations of the Penna ageing model
\cite{penna,progress} suggested that after many centuries, the human genetic 
pool may have acquired so many new mutations, which due to medical intervention
are not removed from the population, that life expectancy and health at old
age no longer improve. 

An example which is only a few decades old is in-vitro fertilisation.
\cite{moll} Normally a preselection of sperm cells happens 
since only an extremely small fraction of sperm cells out of hundreds of 
millions reaches the ovum surface and only one of these groups is allowed to 
enter the cytoplasm. There are some suggestions that an egg can actively 
chose this winner \cite{preselection}. Sperm cells which are slower or 
less viable due to some genetic defects are losers in this competition.
This preselection is
abolished by in-vitro fertilisation and thus may deteriorate the genetic pool.

All these medical interventions are rare: Not everybody takes antibiotics each 
day, and most pregnancies start traditionally and not by in-vitro fertilisation.
Nevertheless, it is also possible that technology of in-vitro 
fertilisation increases the incidence of mutations in the newborns
genomes. Moll at al. \cite{moll} have estimated that the incidence of 
retinoblastoma in children born after in-vitro fertilisation increased 5 
to 7 times comparing with the children born after natural conception. 
Since retinoblastoma is caused by inherited or somatic mutation followed 
by loss of heterogeneity, it is legitimate to ask the question if such a 
small fraction of population with increased mutation load can significantly 
affect the genetic pool and reproduction potential of a population. 
Thus we simulate here that with two-percent probability only the number of 
mutations at birth are enhanced from one to four, and we check how this affects
the {\em overall} population. Can there be detrimental effects far exceeding
the level of a few percent? 

For the simulations we use the Penna model \cite{penna} since it includes ageing
and is widely used \cite{newbook,progress}. Thus we do not repeat here all the
details of that model. We believe that similar results could be found also in 
other ageing models \cite{charlesworth,rose,gavrilov}.

\section{Results}
\subsection{Parameters}

Each individual of the sexual Penna model has two bit-strings of length $L$
each, representing the genome. A mutation changes a bit from
zero to one, or leaves it at one, and is never reversed; at birth one mutation
happens for each bit-string. Age is increased by one bit position at each time
step $t$. Starting from age $R$, each female at each time step selects randomly 
a male of age $\ge R$ and gives birth to $B$ offspring; the offspring dies
immediately with a Verhulst probability $N(t)/K$, where $N$ is the current
population (male plus female) and $K$ is often called the carrying capacity,
varying here from 1000 up to 30 million.  An individual is killed if the
number of mutations expressed at that age reaches a threshold $T$; a mutation
is expressed if in the same positions from one up to the current age both 
bit-strings
have a bit set to one. Thus all mutations are recessive, deleterious and 
inheritable. Before at birth one bit-string from the mother is given on to the 
child and mutated, with a crossover probability $C$ the two maternal bit-strings
are crossed-over at a randomly selected position; the same happens for the 
father. 

Further details on the Penna model are given in many articles and
a few books \cite{newbook}; the influence of recombination is given in
\cite{wroclaw}. We now add new to it that with probability of two
percent the mutation rate increases from one to four for each bit-string
at every birth. We call this case {\em inhomogeneous} and compare with the
standard homogeneous case of always one mutation. (Our inhomogeneity is 
not inherited; if instead we let males give on the property of enhanced mutation
to their male offspring, then the fraction of males with that genetic 
property dies out after about $10^3$ time steps from initially 2 \%. Also,
if both the homogeneous and the inhomogeneous population are simulated together,
with one shared carrying capacity, mostly the inhomogeneous population dies out
for large $K$ and large $C$.) 

Normally we take $R=L/4,\;B=2,\;T=3$ in agreement with numerous previous studies
of the Penna model, but for the transition to complementing bit-strings we use 
\cite{wroclaw} $R=5L/8,\;B=6,\;T=1$ (for $B=5$ in the latter case populations 
became extinct). Throughout we made 20,000 time steps and averaged over the 
second half of the simulation. (Selected simulations up to 1 million time
steps did not change Fig. 1 beyond the large scattering, but increased the 
probability of population extinction.) Except for large $K$ we averaged over 10 
to 1000 samples.

\subsection{Normal case}

Fig. 1 shows the three ratios of the ``homogeneous'' to the ``inhomogeneous''
numbers, for: the total population, the lifespan (more precisely, the life
expectancy at birth), and the reproductive fraction of the population (having
at least the age $R$). We see that these ratios are of the same order as the
two-percent inhomogeneity, i.e. the additional mutations reduce the overall
fitness (as measured by these three quantities). The longer the bit-strings
are, the smaller is the influence of the inhomogeneity, i.e. the closer are
the ratios to unity; this effect is seen even stronger, not shown, for $L > 64$.
The reason is that the longer the bit-strings are, the larger is the total
number of births over the lifetime and thus the closer are the populations $N$ 
to the carrying capacity $K$, not shown, making other influences less
important. Actually the $L$ loci of the Penna model refer to the small
fraction \cite {cebrat} of life-threatening diseases, not to tens of thousands
of less important genes. Fig.2 shows an inbreeding depression for small $K$.

(The error bars in Fig.1 increase with increasing $L$ and decreasing $C$
and are at most about 1/4 percent, from 1000 samples. Later in Fig. 6b, based
on only 100 samples, they can reach 6 percent.) 

While it is plausible that a probability of two percent for an enhanced 
mutation rate reduces the fitness also by a few percent, it is less obvious
that this reduced population is still higher by a few tenths of a percent, Fig.
3, than the one obtained by a mutation rate of 1.06 for all, corresponding to
the weighted average of 1 and 4 mutations used before. (We always make one 
mutation, and then with probability 0.06 another one.) 

Another alternative for the homogeneous case is to have always one mutation,
but to assume infertility in two percent, randomly selected. Even then the
homogeneous case is fitter than the inhomogeneous one, Fig.4.

\subsection{Transition to complementing bit-strings}.

In the above normal case, Darwinian selection of the fittest tries to reduce
the fraction of mutated bits; we call the fraction of bits set to one the
mutation load and the fraction of bit pairs where the two bit-strings differ
(at the same position) the heterozygosity. The mutation load is low, and
the heterozygosity about twice as high.

Instead, for low recombination probabilities $C$ and high minimal reproduction 
ages $R$, and only recessive loci, selection of the fittest leads to 
complementary
bit-strings: At many loci, one bit is set to one and the other to zero. Thus
the load is nearly half, and the heterozygosity nearly 100 percent. Fig. 5
shows this effect \cite{wroclaw}; the higher the population is the lower must
$C$ be to allow for complementarity. The above ratios between the homogeneous
and inhomogeneous cases, Fig. 6, now deviate much stronger from unity at low
crossover rates. 

\section{Discussion}

We put in a two-percent effect, and in most cases got out results also of the 
order of two percent, Fig. 2; this is good. Only at the transition from 
purification to complementarity was the danger enhanced appreciably, Fig. 6. 
Nevertheless, even a one percent reduction in fitness could
be problematic; perhaps {\em Homo neanderthalis} was only a few percent less 
fit than {\em Homo sapiens}, and nevertheless lost out in the competition
after thousands of years.

As an alternative to modern medicine, we do {\em not} advocate eugenics
\cite{pekalski}; and for obvious reasons the present authors also dislike
killing everybody at retirement age. Instead medical research should be aware 
of the long-term dangers and look for better methods reducing these dangers.

\begin{figure}
\begin{center}
\includegraphics[scale=0.32,angle=-90]{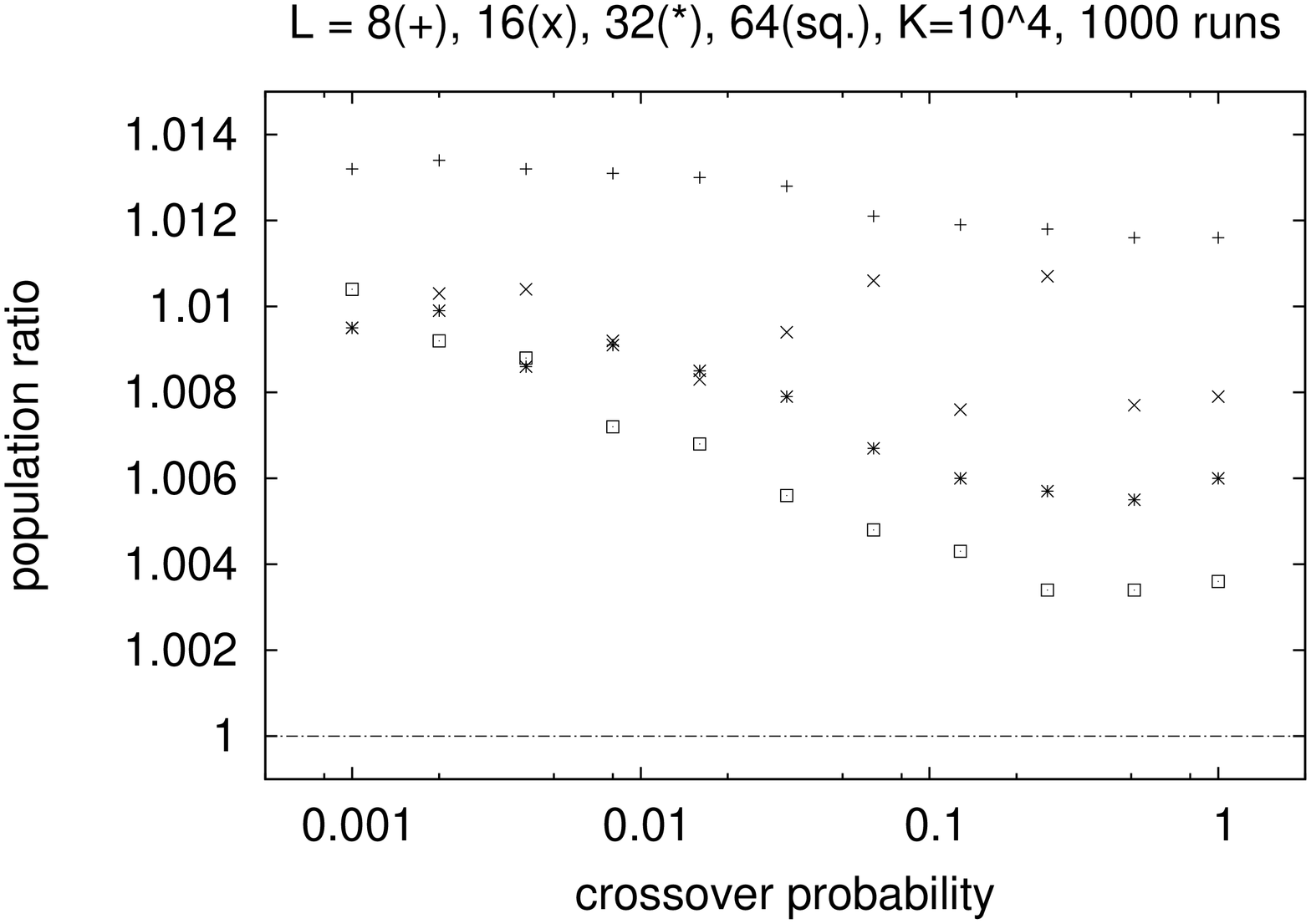}
\includegraphics[scale=0.32,angle=-90]{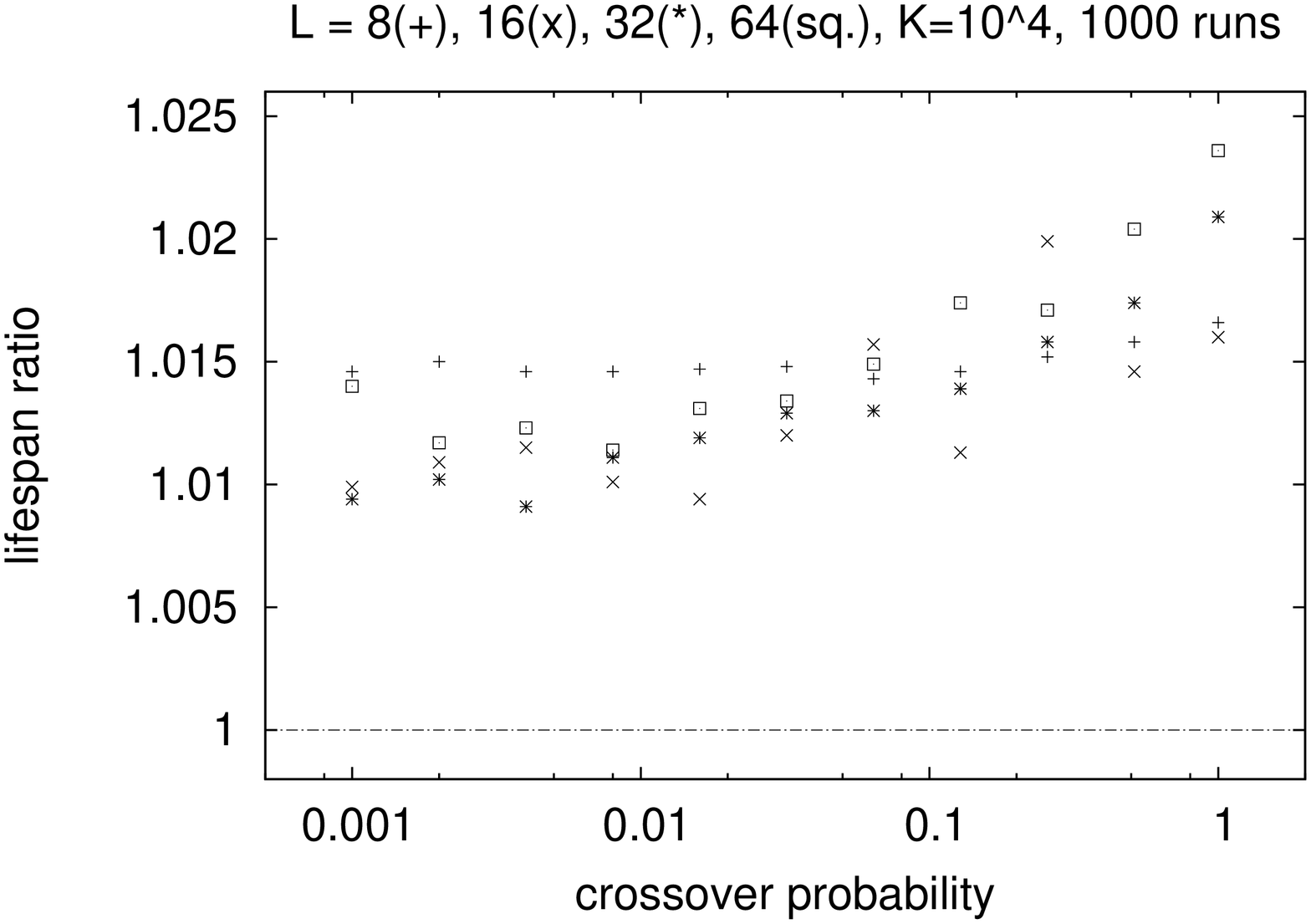}
\includegraphics[scale=0.32,angle=-90]{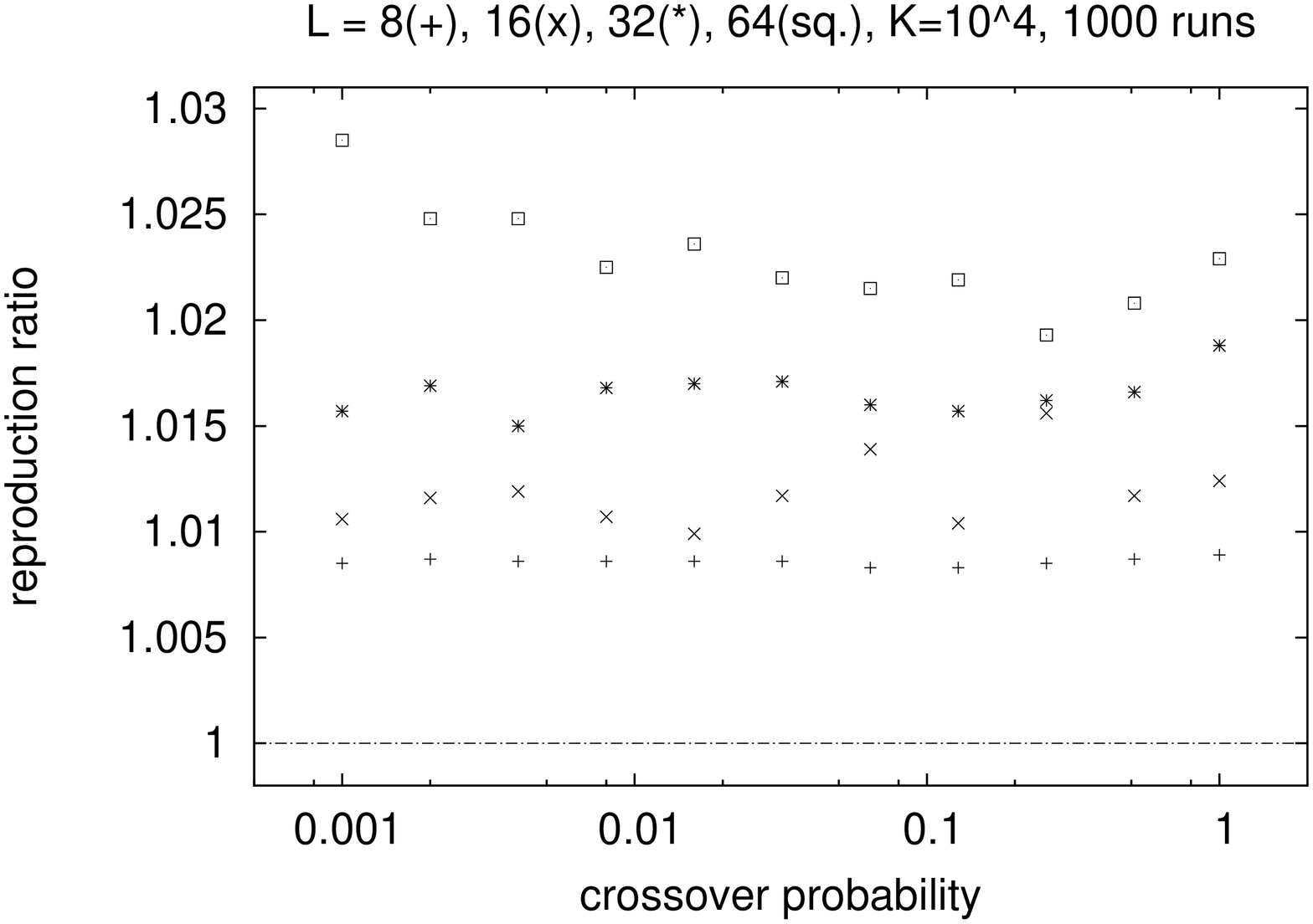}
\end{center}
\caption{
Normal case: The three fitness ratios for a carring capacity of $10^4$.
Similar results were obtained for $K = 10^3, 10^5$ and $10^6$.
}
\end{figure}

\begin{figure}
\begin{center}
\includegraphics[scale=0.50,angle=-90]{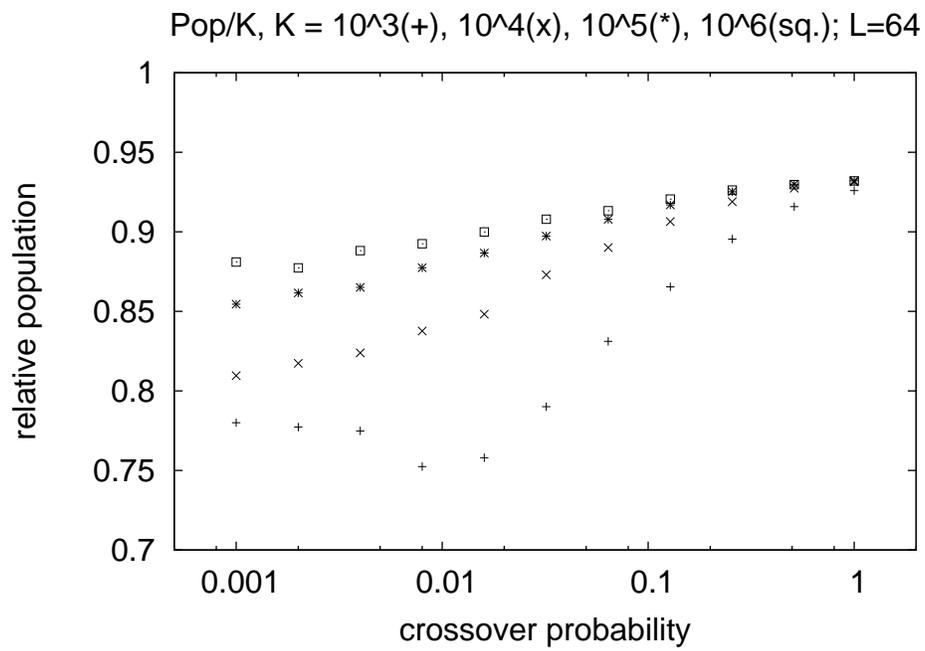}
\end{center}
\caption{
Normal case: Variation of population (normalised by carrying capacity) with $C$.
}
\end{figure}

\begin{figure}
\begin{center}
\includegraphics[scale=0.45,angle=-90]{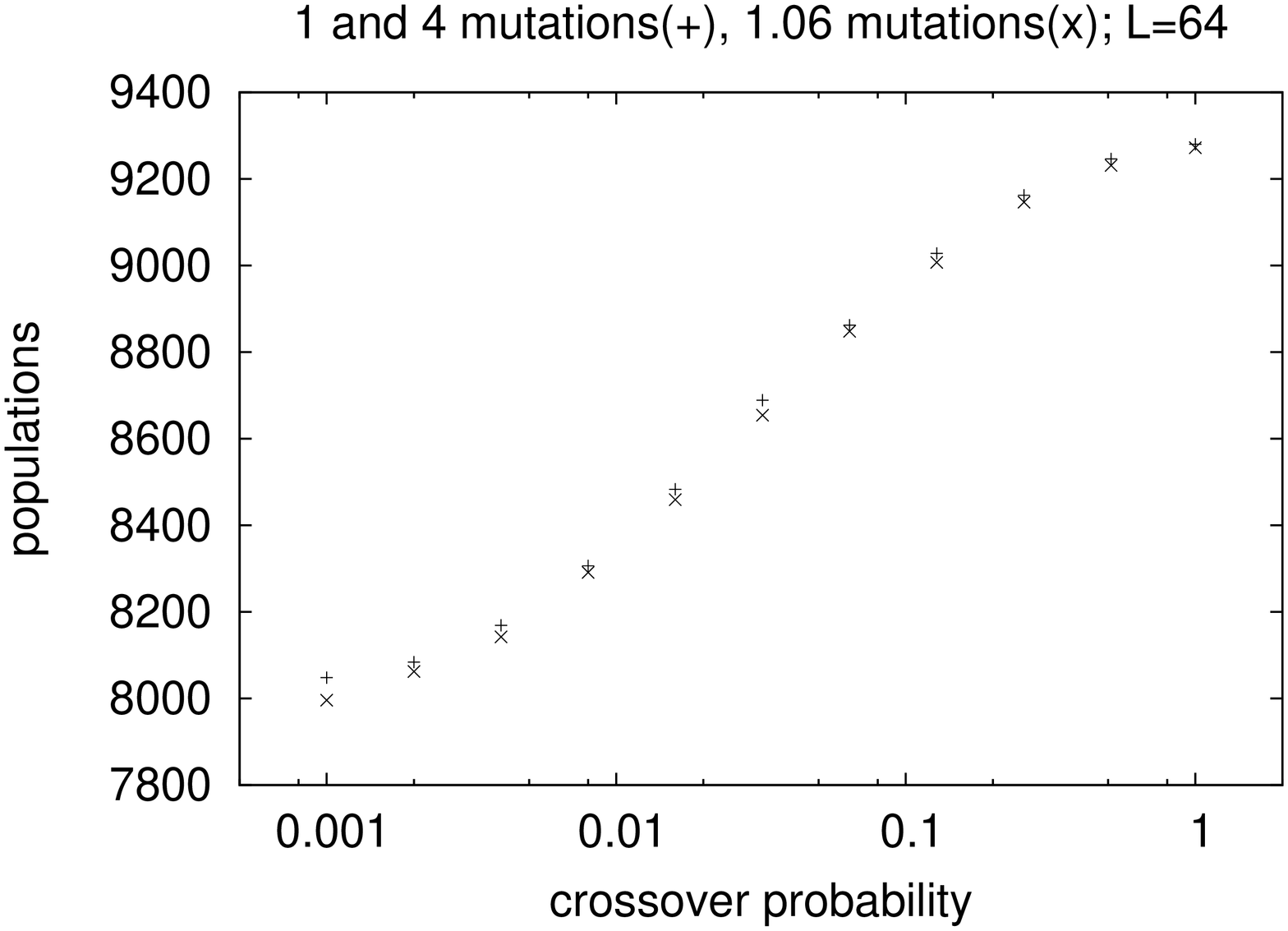}
\includegraphics[scale=0.45,angle=-90]{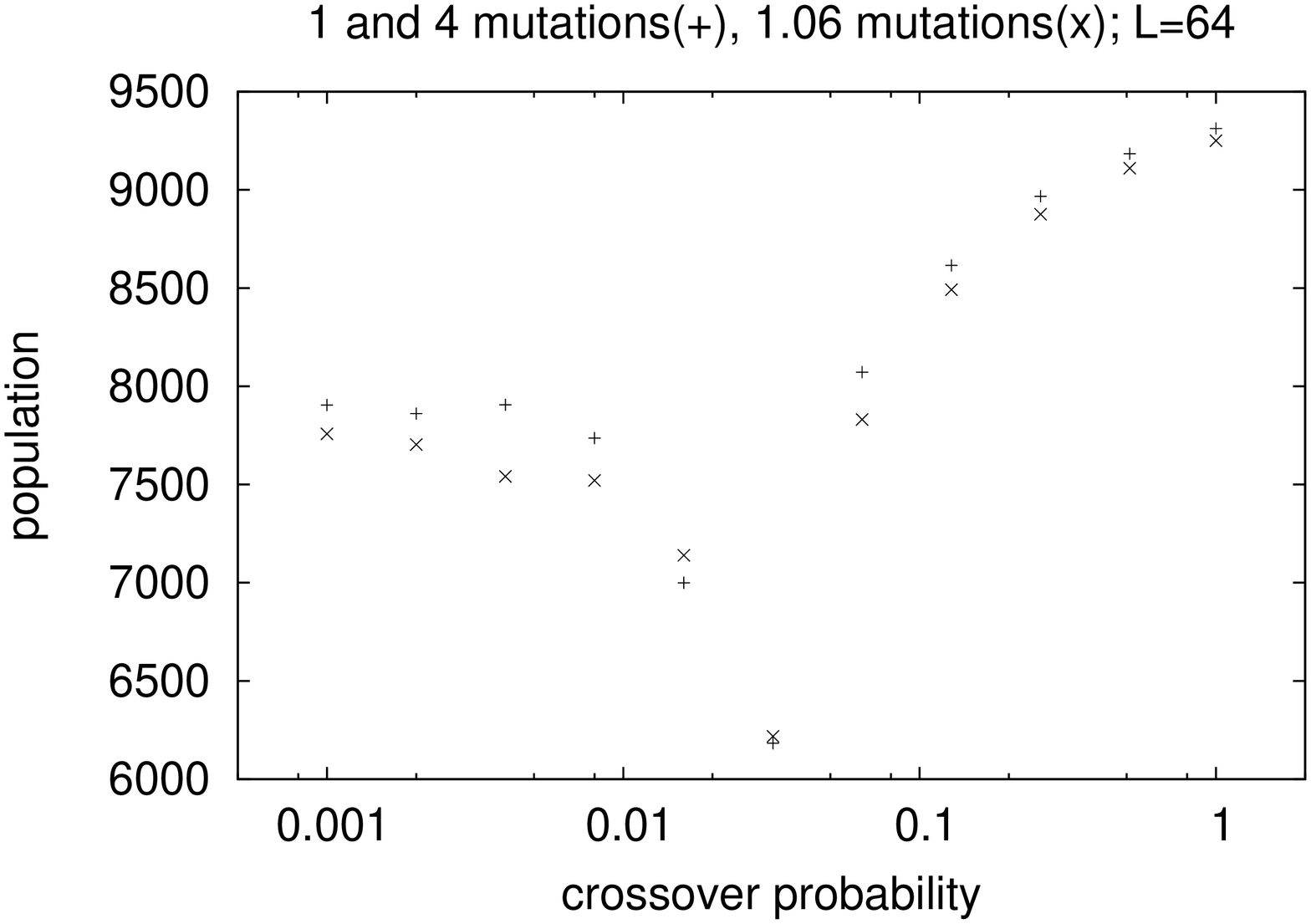}
\end{center}
\caption{
Comparison of inhomogeneous case (higher values) with homogeneous mutation
rate of 1.06 (lower values). The upper figure shows the normal case, the 
lower figure the transition case, $K=10^4$ everywhere.
}
\end{figure}

\begin{figure}
\begin{center}
\includegraphics[scale=0.39,angle=-90]{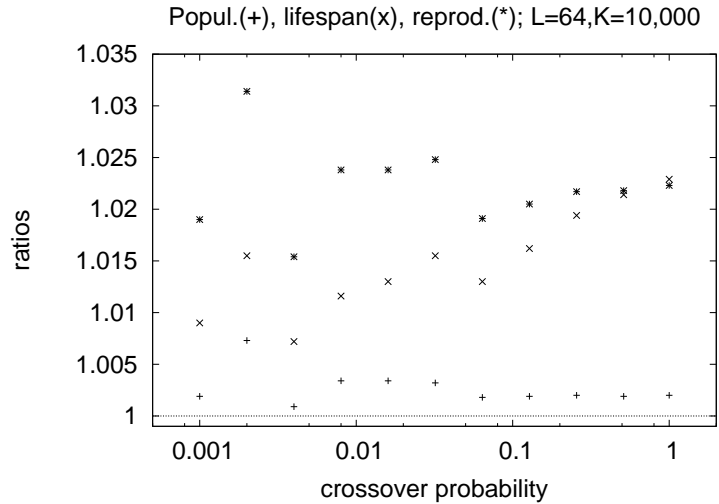}
\end{center}
\caption{
Normal case with two percent infertility in the homogeneous population: 
Variation with $C$ of the three fitness ratios, homogeneous to inhomogeneous.
}
\end{figure}

\begin{figure}
\begin{center}
\includegraphics[scale=0.39,angle=-90]{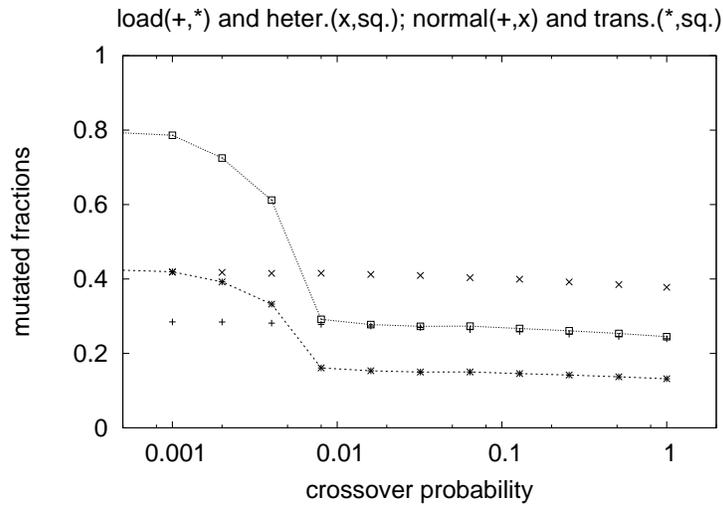}
\end{center}
\caption{
Normal (+,x) and transition (stars, squares) case: Mutation load and 
heterozygosity versus $C$; small $C$ give complementarity
only for the transition case (stars and squares), not the normal case (+ and x).
$K = 10^5$ in both cases.
}
\end{figure}

\begin{figure}
\begin{center}
\includegraphics[scale=0.45,angle=-90]{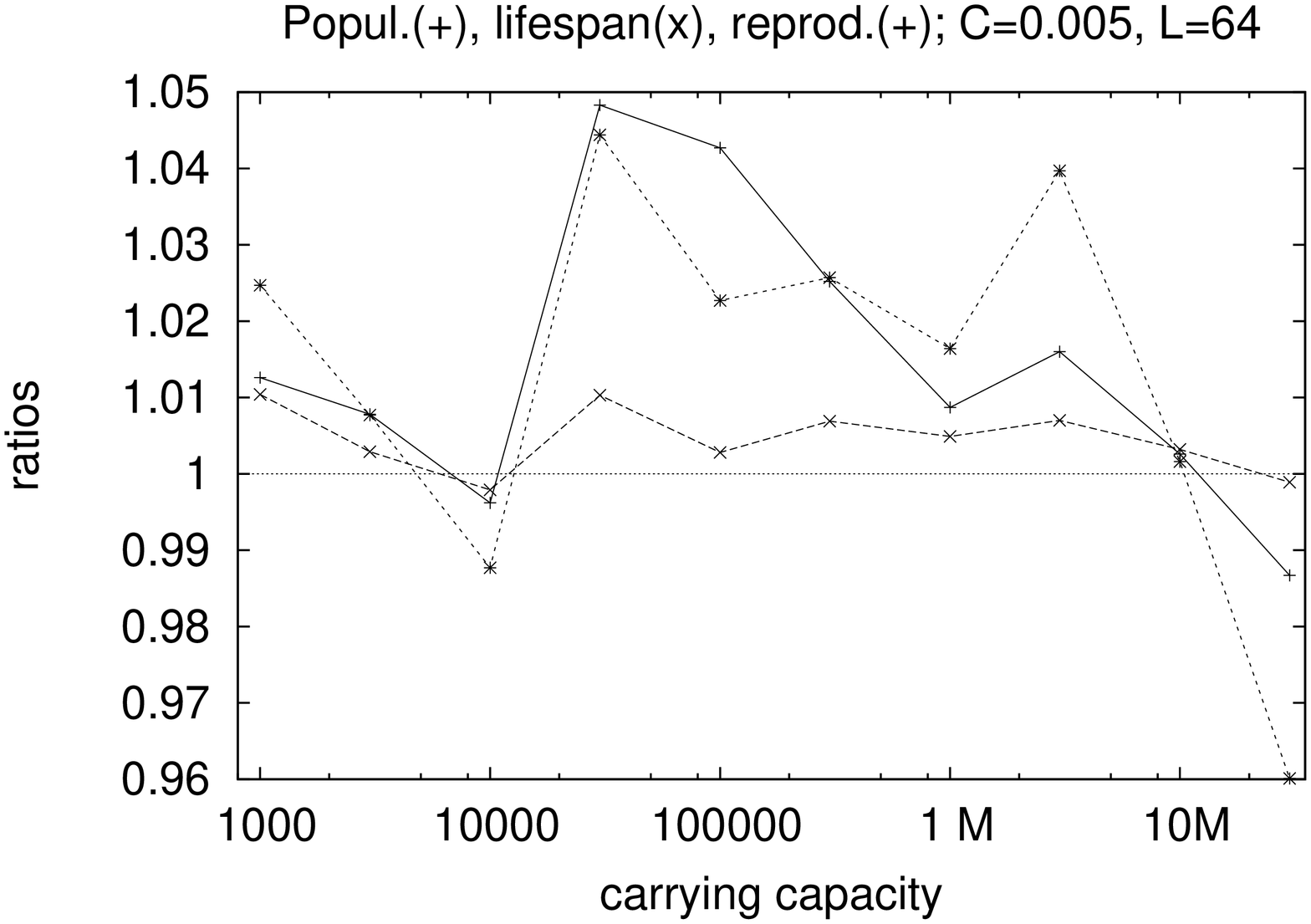}
\includegraphics[scale=0.45,angle=-90]{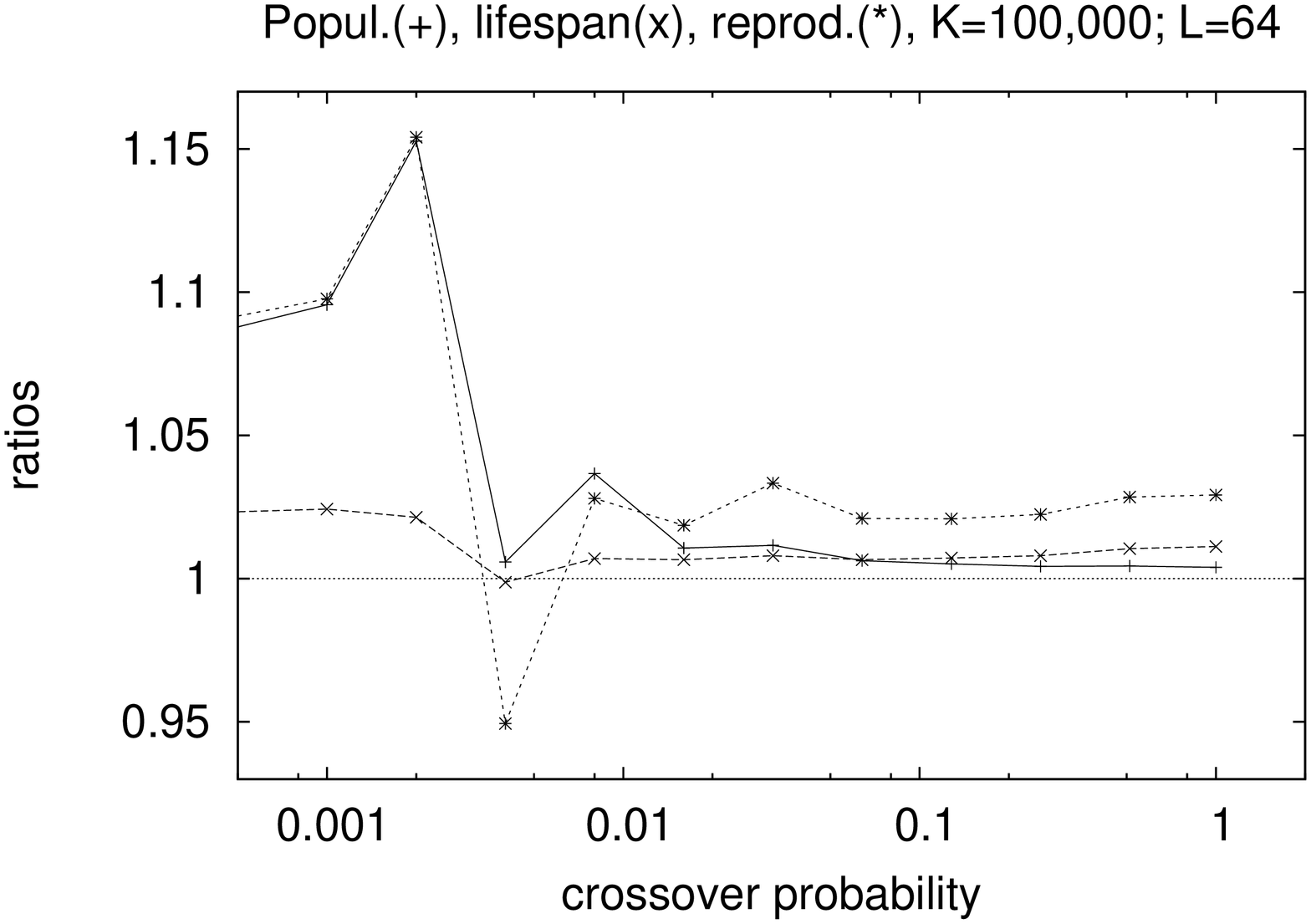}
\end{center}
\caption{Transition case: Fitness ratios versus $K$ and versus $C$ (ratio
of homogeneous case with one mutation, to inhomogeneous case with one and
four mutations).
}
\end{figure}
\end{document}